\documentclass[aps,12pt]{revtex4}
\usepackage{amsmath}
\usepackage{amsfonts}
\usepackage{graphicx}
\usepackage{epstopdf}
\usepackage{appendix}
\usepackage{hyperref}
\hypersetup{
    colorlinks,
    citecolor=blue,
    filecolor=blue,
    linkcolor=blue,
    urlcolor=blue
}
\begin{document}

\title{On the quadruple Beltrami fields in thermally relativistic
electron-positron-ion plasma}
\author{Usman Shazad*}
\author{M Iqbal}
\affiliation{Department of Physics, University of Engineering and Technology, Lahore
54890, Pakistan}
\email{usmangondle@gmail.com}

\begin{abstract}
A thermally relativistic electron-positron-ion (EPI) plasma self-organizes
into a quadruple Beltrami (QB) field. The QB field, which is the combination
of four Beltrami fields, is described by four scale parameters. These scale
parameters are often either real or both real and complex in nature. The
values of the scale parameters are determined by Beltrami parameters,
relativistic temperatures, and the densities of plasma species. It is
demonstrated that all the scale parameters become real at higher
relativistic temperatures and ion densities, which naturally lead to
paramagnetic structures. It is also shown that the scale separation
in the QB state provides the possibility of field and flow generation in
such thermally relativistic plasmas. The present study may have implications for
space, astrophysical and laboratory plasmas.
\end{abstract}

\keywords{Beltrami fields, relativistic plasma, self-organization,
multiscale structures, electron-positron-ion}
\maketitle

\section{Introduction}, \label{s1}

Magnetic self-organization, commonly referred to as plasma relaxation, is
the process of minimizing plasma energy under helicity constraints. The
Beltrami fields are typically used to characterize these relaxed states. A
Beltrami field is an eigenvector of the curl operator. For instance, when
the magnetic energy of an ideal MHD plasma under magnetic helicity
constraint is minimized, the relaxed state is\ a single Beltrami state,
which is expressed as $\mathbf{\nabla }\times \mathbf{B}=\mu \mathbf{B}$,
where $\mu $ is constant and eigenvalue of the curl operator. In the single
Beltrami state, the current density is aligned to the magnetic field, so the
relaxed state is force-free and flow-less \cite%
{Woltjer1958,Taylor1974,Taylor1986}.

Plasmas in nature or a laboratory aren't force-free and flow-less, hence the
single fluid MHD theory was extended to account for novel relaxed states
with significant plasma flow and pressure gradients. Multifluid plasma
self-organization models suggest a non-force-free relaxed state. The
self-organized state of multifluid plasmas can be described as a combination
of multiple Beltrami fields. For example, the relaxed state in Hall MHD
represents the combination of two Beltrami fields, which is referred to as a
double Beltrami (DB) state \cite{Steinhauer1997,Mahajan1998,Steinhauer2002}.
On the other hand, the inclusion of inertial effects results in a relaxed
state that is composed of three Beltrami fields and is referred to as a
triple Beltrami (TB) state. In comparison to MHD relaxation, relaxed
multifluid plasmas have substantial flow and pressure gradients. When
compared to single fluid MHD relaxation, which has only one macroscopic
scale, the self-organized state of two or multicomponent plasmas is
characterized by multiscale structures (both macroscopic and microscopic
structures) \cite%
{Bhattacharyya2003,Iqbal2012POP,Iqbal2013APL,Mahajan2015,Shafa2022}. The
multi-Beltrami relaxed states in plasmas allow for the formation of very
novel and complex structures, which are applied to the study of a wide range
of physical phenomena that occur in the laboratory and in space. For
instance, high beta relaxed states in tokamak \cite%
{Mahajan2000,Yoshida2001,Guzdar2005}, formation and heating of solar corona 
\cite{Mahajan2001,Mahajan2002}, eruptive events in space plasma \cite%
{Ohsaki2002,Kagan2010}, acceleration of plasma flows \cite{Mahajan2006},
dynamo/reverse dynamo mechanisms \cite{Mininni2002,Mahajan2005}, turbulence 
\cite{Hamdi2016,Mahajan2020} and etc.

The goal of this study is to explore the relaxed states of a thermally
relativistic electron, positron and ion (EPI) plasma. The EPI plasmas can be
found in a number of different astrophysical environments and can also be
created in a laboratory. These astrophysical environments include the early
universe \cite{Holcomb1989}, pulsar magnetospheres \cite%
{Sturrock1971,Ruderman1975,Karpman1975,Melrose1978,Michel1982}, accretion
disks around black holes \cite{Liang1988}, active galactic nuclei (AGN) \cite%
{Begelman1984}, gamma-ray bursts \cite{Furlanetto2002}, supernova remnants 
\cite{Martin2010} and solar flares \cite{Murphy2005}. In the laboratory it
is also possible to produce EPI plasmas by injecting positrons into the
electron ion system and through ultra-intense laser matter interactions \cite%
{Surko1986,Greaves1997,Helander2003,Fajans2020,Sarri2015,Peebles2021}. The
EPI plasmas in these astrophysical objects and laboratory experiments are
usually in the relativistic regime. Plasma is relativistic when its fluid
velocity approaches the speed of light or when its average kinetic energy
surpasses its rest mass energy.

In recent years, many researchers have studied self-organization in
relativistic plasmas. A study by Iqbal et al. has shown that the relaxed
state of a relativistically hot EP plasma is a TB state. It has been
demonstrated that at higher thermal energy, all of the scale parameters are
real, and the size of one of the structures grows larger with an increase in
the amount of thermal energy \cite{Iqbal2008}. The relaxed state is likewise
a TB state when all of the inertial effects of a relativistically hot
three-component EPI plasma are taken into consideration. It has been shown
that a rise in relativistic temperature affects one self-organized structure
more than the other two structures. At higher relativistic temperatures, two
separate structures eventually become a single one. This demonstrates that
diamagnetic plasma structures in such relativistic hot plasmas are possible 
\cite{Iqbal2012,Iqbal2013}. In another study, the variational approach was
used to obtain a relaxed state for a fully relativistic plasma (one in which
the fluid velocity as well as the thermal energy are both
relativistic). This state was used to replicate the striped wind phenomena
that occur in the pulsar nebula \cite{Pino2010}. For~a relativistically hot
EPI plasma that contains static ions, the self-organized state is also a TB
state. The parametric study of the relaxed state reveals that when the
thermal energy and ion density increase, the scale separation grows, and the
diamagnetic structures eventually transform into the paramagnetic structures 
\cite{Usman2021}.

In the past few years, the field of plasma relaxation has been extended to
include relativistic degenerate plasmas. It has been investigated that a
Beltrami-Bernoulli equilibrium state with electron degeneracy pressure is
also possible. Even in a zero-temperature plasma, a Beltrami-Bernoulli
equilibrium state can be obtained due to electron degeneracy. These states
are being investigated in order to better depict new pathways for the
transformation of energy, such as the conversion of degeneracy energy to
fluid kinetic energy \cite{Berezhiani2015}. Shatashvili et al. have further
explored the relaxed states of such relativistic degenerate plasmas to
understand the impact of relativistic degeneracy on the multiscale relaxed
state structures. It has been studied that a three-component plasma with
relativistic degenerate electrons and positrons and mobile classical ions is
a quadruple Beltrami (QB) state. In the same way, a QB state can also be
reached with a three-component plasma made up of mobile classical ions and
electrons with two temperatures (relativistic degenerate and classical hot).
The investigation of these QB states has shed light on the role that
degeneracy-induced relativistic temperature has on the formation of
multiscale structures \cite{Shatashvili2016,Shatashvili2019}. Plasma
relaxation is also the focus of a great deal of research that takes into
consideration the implications of general relativity as such relativistic plasmas
are also encountered in the vicinity of black holes \cite%
{Bhattacharjee2015,Bhattacharjee2019,Asenjo2019,Bhattacharjee2020}.

In the present study which is extension of our previous work \cite{Usman2021}%
, we consider a three-component thermally relativistic EPI plasma. In a
thermally relativistic plasma, the thermal energy of plasma species is on
the order of or greater than their rest mass energy, whereas the
fluid velocity of the plasma species is considered to be non-relativistic. 
In recent years, a lot of research has been done on such EPI plasmas
with thermal relativistic effects \cite%
{Pokhotelov2001,Eliasson2005,Tsintsadze2013,Rozina2014,Lu2015,Petropoulou2019}.
In our plasma model, the relativistic temperatures of electrons and
positrons are assumed to be equal, whereas the relativistic temperature of
ions is taken differently and considered mildly relativistic. Such thermally
relativistic two-temperature plasmas with sub-relativistic ions are created
in low density and high temperature astrophysical systems due to
relativistic radiative turbulence. The some examples of such astrophysical
systems are hot accretion flows, the galactic center, M87 and pulsar
magnetospheres \cite{Zhdankin2019,Zhdankin2021}. It is shown that the
relaxed state is the QB state. The effect of plasma parameters on the
relaxed state has been investigated. For given values of the Beltrami
parameters, the density and thermal energy of pair species have been
demonstrated to be able to convert diamagnetic structures into paramagnetic
ones and vice versa. The study also highlights the impact of scale
separation in QB state in the context of dynamo and reverse dynamo mechanisms.

The structure of the paper is as follows: The model equations and the QB
relaxed state equation is derived in Sec. \ref{s2}. The characteristics of the
scale parameters are discussed in Sec. \ref{s3}. The following section provides
an analytical solution for the QB field and flow in simple slab geometry and
investigates the impact of plasma parameters and scale separation on relaxed
state structures. The summary of the current study can be found in Sec. \ref{s5}.

\section{Model equations and QB fields}, \label{s2}

In this study, we consider a three-component incompressible, collisionless,
quasi-neutral and magnetized plasma whose components are electrons,
positrons and ions (in model equations we will use $\alpha $ for plasma
species, $e$--electron, $p$--positron and $i$--ion). All of the plasma
species are dynamic and only thermally relativistic. A plasma is said to be
relativistic when the directed fluid velocity approaches the speed of light
or the thermal energy of plasma particles is the order of or equal to the
rest mass energy. Both approaches to relativity emerge in astrophysical and
laboratory plasmas. By following Ref. \cite{Berezhiani1995}, if the velocity
distribution of $\alpha $ plasma species is local relativistic Maxwellian,
the relativistic equation of motion can be expressed as:%
\begin{equation}
\frac{\partial \mathbf{P}_{\alpha }}{\partial t}+m_{o\alpha }c^{2}\mathbf{%
\nabla }\left( \gamma _{\alpha }G_{\alpha }\right) =q_{\alpha }\mathbf{E}+%
\mathbf{V}_{\alpha }\times \left( \mathbf{P}_{\alpha }+\frac{q_{\alpha }}{c}%
\mathbf{B}\right) ,  \label{REM}
\end{equation}%
where $\mathbf{P}_{\alpha }=\gamma _{\alpha }G_{\alpha }m_{0\alpha }\mathbf{V%
}_{\alpha }$, $c$, $\mathbf{V}_{\alpha }$, $\gamma _{\alpha }$, $G_{\alpha }$%
, $m_{0\alpha }$ and $q_{\alpha }$ are relativistic momentum, speed of
light, plasma velocity, Lorentz factor ($1/\sqrt{1-V_{\alpha }^{2}/c^{2}}$),
relativistic temperature, rest mass and electric charge respectively. The $%
\mathbf{E}$ and $\mathbf{B}$ are electric and magnetic fields which are
related to scalar electric potential $\phi $ and vector magnetic potential $%
\mathbf{A}$ by the following relations: $\mathbf{E}=\mathbf{-\nabla }\phi
-c^{-1}\partial \mathbf{A/\partial }t$ and $\mathbf{B}=\mathbf{\nabla }%
\times \mathbf{A}$. In above equation the streaming relativistic effects
accounted through $\gamma _{\alpha }$ while the thermal relativistic effects
appear through the factor $G_{\alpha }=K_{3}(1/z_{\alpha
})/K_{2}(1/z_{\alpha })$, where $K_{2}$ and $K_{3}$ are the modified Bessel
functions and $z_{\alpha }=T_{\alpha }/m_{0\alpha }c^{2}$, where $T_{\alpha
} $ is proper temperature of the plasma species. For non-relativistic case $%
z_{\alpha }<<1,$ and the factor $G_{\alpha }$ can approximately be taken as $%
G_{\alpha }\simeq 1+5z_{\alpha }/2$, whereas for highly relativistic plasma
species $z_{\alpha }>>1$ and the relativistic temperature $G_{\alpha }$ can
be approximated as $G_{\alpha }\simeq 4z_{\alpha }$. It is essential to
point out that the preceding equation of motion is supplemented with the
following equation of state:%
\begin{equation}
\frac{n_{\alpha }}{\gamma _{\alpha }}\frac{z_{\alpha }}{K_{2}\left(
z_{\alpha }\right) }\exp \left( -z_{\alpha }G_{\alpha }\right) =\text{
constant.}
\end{equation}
Following Shatashvili et al. \cite{Shatashvili2016}, we'll start
with macroscopic evolution equations for multifluid relativistic hot plasma
to derive the QB state. The curl of these equations will yield vorticity evolution
equations. The steady state solution of these vorticity evolution equations
provides three Beltrami conditions. The Beltrami conditions basically show
the alignment of generalized vorticity with the respective flow. To couple
the steady state dynamics of plasma species given by the Beltrami condition,
we will employ Ampere's law. By solving these three Beltrami conditions
along with Ampere's law, we will obtain a QB relaxed state equation.

Since we will only be considering plasma that is thermally relativistic, the
flow velocity of plasma species will be non-relativistic and hence \ $\gamma
_{\alpha }\approx 1$. Furthermore, we assume that electrons and positrons
have the same relativistic temperature\ ($G_{e}=G_{p}=G$) whereas ions have
a different relativistic temperature ($G_{i}$). The quasi-neutrality
condition for the plasma system is $N_{i}+N_{p}=1$, where $N_{i}$ and $N_{p}$
are $n_{0i}/n_{0e}$ and $n_{0p}/n_{0e}$ respectively, in which $n_{0e}$, $%
n_{0p}$ and $n_{0i}\ $are the number densities of electron, positron and ion
species in rest frame of reference. The macroscopic evolution equations of
thermally relativistic two temperature EPI\ plasma in normalized form (for
normalization of length, plasma species velocity, temperature and magnetic
field we have used electron skin depth $\lambda _{e}=\sqrt{m_{0e}c^{2}/4\pi
n_{0e}e^{2}}$, Alfv\'{e}n speed $v_{A}=B_{0}/\sqrt{4\pi m_{0e}n_{0e}}$,
electron plasma frequency $\omega _{pe}=\sqrt{4\pi n_{e}e^{2}/m_{0}}$, rest
mass energy of electron $m_{0e}c^{2}$ and some arbitrary value of magnetic
field $B_{0}$, respectively) can be expressed as%
\begin{equation}
\frac{\partial \mathbf{P}_{\alpha }}{\partial t}=\mathbf{V}_{\alpha }\times 
\mathbf{\Omega }_{\alpha }-\mathbf{\nabla }\psi _{\alpha },  \label{EME6}
\end{equation}%
where $\mathbf{P}_{e}=G\mathbf{V}_{e}-\mathbf{A}$, $\mathbf{P}_{p}=G\mathbf{V%
}_{p}+\mathbf{A}$, $\mathbf{P}_{i}=G_{i}\mathbf{V}_{i}+M\mathbf{A}$,\textbf{%
\ }$\mathbf{\Omega }_{e}=\mathbf{\nabla \times P}_{e}$,\textbf{\ }$\mathbf{%
\Omega }_{p}=\mathbf{\nabla \times P}_{p}$,\textbf{\ }$\mathbf{\Omega }_{i}=%
\mathbf{\nabla \times P}_{i}$, $\psi _{e}=-\phi -G$, $\psi _{p}=\phi -G$ and 
$\psi _{i}=M\phi -G_{i}$ and $M=m_{0e}/m_{0i}$ in which $m_{0e}$ and $m_{0i}$
are rest masses of electron and ion. In above equations of motion $\mathbf{P}%
_{\alpha }$ and $\mathbf{\Omega }_{\alpha }$ are termed as
canonical/generalized momentum and vorticity respectively. To couple the
dynamics of plasma species and close the system of equations, Ampere's law
is adopted. For this plasma system, Ampere's law in dimensionless form is%
\begin{equation}
\mathbf{\nabla }\times \mathbf{B}=N_{p}\mathbf{V}_{p}+N_{i}\mathbf{V}_{i}-%
\mathbf{V}_{e}.  \label{AL6}
\end{equation}

To obtain the vorticity evolution equations we take curl of the momentum
equations (\ref{EME6}), the following vorticity evolution equations are
obtained%
\begin{equation}
\frac{\partial \mathbf{\Omega }_{\alpha }}{\partial t}=\mathbf{\nabla }%
\times \lbrack \mathbf{V}_{\alpha }\times \mathbf{\Omega }_{\alpha }].
\label{VEE6}
\end{equation}

The steady-state solution of equation (\ref{VEE6}) with the condition $%
\mathbf{V}_{\alpha }\times \mathbf{\Omega }_{\alpha }=0$, defines the
multi-Beltrami equilibrium state provided that the gradient forces $\mathbf{%
\nabla }\psi _{\alpha }$ are separately constrained to zero. The latter
provides the generalized Bernoulli conditions, which are necessary for
closure, although they are not important for the analysis described in this
article. This solution yields three Beltrami conditions ($\mathbf{V}_{\alpha
}\parallel \mathbf{\Omega }_{\alpha }$) for electrons, positrons and ions.
These Beltrami conditions are given by%
\begin{equation}
\mathbf{\nabla }\times G\mathbf{V}_{e}-\mathbf{B}=aG\mathbf{V}_{e},
\label{BCE6}
\end{equation}%
\begin{equation}
\mathbf{\nabla }\times G\mathbf{V}_{p}+\mathbf{B}=bG\mathbf{V}_{p},
\label{BCP6}
\end{equation}%
\begin{equation}
\mathbf{\nabla }\times G_{i}\mathbf{V}_{i}+M\mathbf{B}=cG_{i}\mathbf{V}_{i},
\label{BCI6}
\end{equation}%
where $a$, $b$ and $c$ are the Beltrami parameters.\textbf{\ }The Beltrami
parameters are ratios of generalized vorticities to their respective flows.%
\textbf{\ }The equations (\ref{AL6} and \ref{BCE6}-\ref{BCI6}) will be
employed to derive a relaxed state. By solving these equations
simultaneously, the expression for ion species velocity $\boldsymbol{V}_{i}$
in terms of magnetic field can be expressed as%
\begin{equation}
\mathbf{V}_{i}=i_{4}(\mathbf{\nabla }\times )^{3}\mathbf{B}-i_{3}(\mathbf{%
\nabla }\times )^{2}\mathbf{B}+i_{2}\mathbf{\nabla }\times \mathbf{B}-i_{1}%
\mathbf{B},  \label{IV6}
\end{equation}%
where $i_{4}=\alpha _{3}^{-1}$, $i_{3}=\left( a+b\right) \alpha _{3}^{-1}$, $%
i_{2}=\left( ab+\alpha _{1}\right) \alpha _{3}^{-1}$, $i_{1}=\left( \alpha
_{2}+\alpha _{1}b\right) \alpha _{3}^{-1}$, $\alpha _{1}=\left(
1+N_{p}\right) G^{-1}+MN_{i}G_{i}^{-1}$, $\alpha _{2}=N_{p}\left( a-b\right)
G^{-1}+MN_{i}\left( a-c\right) G_{i}^{-1}$ and $\alpha _{3}=N_{i}(a-c)(b-c)$%
. Similarly the expression for $\boldsymbol{V}_{p}$ in terms of $\mathbf{B}$
can be written as%
\begin{equation}
\mathbf{V}_{p}=p_{4}(\mathbf{\nabla }\times )^{3}\mathbf{B}-p_{3}(\mathbf{%
\nabla }\times )^{2}\mathbf{B}+p_{2}\mathbf{\nabla }\times \mathbf{B}-p_{1}%
\mathbf{B},  \label{PV6}
\end{equation}%
where $p_{4}=i_{4}N_{i}p(a-c)$, $p_{3}=i_{3}N_{i}p(a-c)-p$, $%
p_{2}=i_{2}N_{i}p(a-c)\mathbf{-}ap$, $p_{1}=b_{1}N_{i}p(a-c)\mathbf{-}\alpha
_{1}p$ and $p=N_{p}^{-1}(b-a)^{-1}$. By substituting the values of $%
\boldsymbol{V}_{i}$, $\boldsymbol{V}_{p}$ from equations (\ref{IV6}-\ref{PV6})
into equation (\ref{AL6}), the expression for $\boldsymbol{V}_{e}$ is%
\begin{equation}
\mathbf{V}_{e}=e_{4}(\mathbf{\nabla }\times )^{3}\mathbf{B}-e_{3}(\mathbf{%
\nabla }\times )^{2}\mathbf{B}+e_{2}\mathbf{\nabla }\times \mathbf{B}-e_{1}%
\mathbf{B},  \label{EV6}
\end{equation}%
where $e_{4}=N_{p}p_{4}+N_{i}i_{4}$, $e_{3}=N_{p}p_{3}+N_{i}i_{3}$, $%
e_{2}=N_{p}p_{2}+N_{i}i_{2}-1$ and $e_{1}=N_{p}p_{1}+N_{i}i_{1}$. To view
the behavior of composite flow of plasma in relaxed state we find the
relation for composite velocity $\boldsymbol{V}$ given by%
\begin{equation}
\boldsymbol{V}=\frac{M\boldsymbol{V}_{e}+MN_{p}\boldsymbol{V}_{p}+N_{i}%
\boldsymbol{V}_{i}}{M+MN_{p}+N_{i}}\mathbf{.}
\end{equation}%
By plugging in the values of $\boldsymbol{V}_{i}$, $\boldsymbol{V}_{p}$ and $%
\boldsymbol{V}_{e}$ from equations (\ref{IV6}-\ref{EV6}) in above relation,
the relation for $\mathbf{V}$ becomes%
\begin{equation}
\mathbf{V}=f_{4}(\mathbf{\nabla }\times )^{3}\mathbf{B}-f_{3}(\mathbf{\nabla 
}\times )^{2}\mathbf{B}+f_{2}\mathbf{\nabla }\times \mathbf{B}-f_{1}\mathbf{B%
},  \label{CFV6}
\end{equation}%
where $f_{j=1,2,3,4}=f\left( N_{i}i_{j}+MN_{p}p_{j}+Me_{j}\right) $
and $f=\left( M +MN_{p}+N_{i}\right) ^{-1}$. To obtain a relaxed state
equation for magnetic field, we substitute the value of $\mathbf{V}_{i}$
from equation (\ref{IV6}) in equation (\ref{BCI6}) which leads to the
following equation%
\begin{equation}
(\mathbf{\nabla }\times )^{4}\mathbf{B}-k_{4}(\mathbf{\nabla }\times )^{3}%
\mathbf{B}+k_{3}(\mathbf{\nabla }\times )^{2}\mathbf{B}-k_{2}\mathbf{\nabla }%
\times \mathbf{B}+k_{1}\mathbf{B}=0,  \label{QB6}
\end{equation}%
where $k_{4}=a+b+c$, $k_{3}=ab+ac+bc+G^{-1}(1+N_{p} +G_{i}^{-1}MN_{i}$), $%
k_{2}=G^{-1}(b+c+aN_{p}+cN_{p}) +abc+G_{i}^{-1}MN_{i}(a+b)$ and $%
k_{1}=cG^{-1}(b+aN_{p}) +abG_{i}^{-1}MN_{i}$. The relaxed state equation (%
\ref{QB6}), is QB equation which can be expressed as superposition of four
single Beltrami fields. The QB state in thermally relativistic two
temperature EPI plasma is the consequence of taking into account inertia of
all the plasma species and three distinct Beltrami parameters for each
plasma species. It is also very important to note that the expressions for
plasma species velocities and composite velocity show strong magnetofluid
coupling and all the vector fields ($\mathbf{V}_{e}$, $\mathbf{V}_{p}$, $%
\mathbf{V}_{i}$ and $\mathbf{B}$) are QB fields for this plasma model.

In this plasma model lower index Beltrami states (DB and TB) can also be
obtained by adjusting the Beltrami parameters. For instance, a TB relaxed
state emerges when the ratios of generalized vorticity to the flow of any
two plasma species are equal to one another (have the same values of
Beltrami parameters) \cite{Iqbal2013}. Similar is the case when the flow
vorticity of any two plasma species is aligned to magnetic field, the
relaxed state comes out to be TB state. If the flows of plasma species are
adjusted in such a way that the ratios of generalized vorticities to flows
become equal for all of the plasma species, one can obtain a DB state.

In 2008, Mahajan demonstrated theoretically that a perfect
diamagnetic state is also possible in a magnetized classical plasma. To
achieve this diamagnetic relaxed state, which he termed as
"classical perfect diamagnetism (CPD)" in classical single-fluid
plasma, flow vorticity was aligned with the magnetic field. He also proposed
that by experimenting with different methods of injecting particle beams
into the ambient plasma, it is possible to create an infinitesimally small
helicity of plasma species \cite{Mahajan2008}. In order to get a glimpse of
CPD in thermally relativistic EPI, we assume that the flow vorticities of
all the plasma species become parallel to the magnetic field. This is a
special class of steady state solution to the vorticity evolution equations.
The vanishing of generalized helicity of all the plasma species is the
singular limit for this plasma system, which is only feasible when
electromagnetic, kinematic and thermal forces are all in balance. So from
equations (\ref{AL6} and \ref{BCE6}-\ref{BCI6}), under above assumption we
obtain the following equation for CPD%
\begin{equation}
\mathbf{\nabla }\times \mathbf{\nabla }\times \mathbf{B}+k\mathbf{B}=0,
\label{CPD}
\end{equation}%
where $k=G^{-1}(1+N_{p})+MN_{i}G_{i}^{-1}$. It is also
clear from equation (\ref{CPD}) that relativistic temperatures and plasma
species densities can also influence the strength of diamagnetism in the
relaxed state. More recently, Asenjo and Mahajan studied similar perfect
diamagnetic~states in a cosmic plasma consisting of electrons and positrons
in the radiation epoch of the early universe. They proposed that for the
expanding cosmological plasmas in a curved space-time, the electromagnetic,
kinematic, and thermal forces can come to balance, resulting in the
disappearance of the generalized helicities of plasma species, and as a
result, there is a diamagnetic trend in the magnetic field \cite{Asenjo2019}.

\section{Characteristics of scale parameters}, \label{s3}

The commutative nature of curl operator dictate us that the QB equation (\ref%
{QB6}) can also be written as%
\begin{equation*}
\left( \text{curl}-\mu _{1}\right) \left( \text{curl}-\mu _{2}\right) \left( 
\text{curl}-\mu _{3}\right) \left( \text{curl}-\mu _{4}\right) \mathbf{B}=0,
\end{equation*}%
where $\mu _{1}$, $\mu _{2}$, $\mu _{3}$, and $\mu _{4}$ are the eigenvalues
of the curl operator and also called scale parameters \cite{Yoshida1990}.
The scale parameters are basically the ratio of current density to magnetic
field, and they show twisting or shearing of the magnetic field The
dimensions of these scale parameters are inverse of length, so the size of
self-organized structures is determined by them whereas the behavior of
relaxed state (paramagnetic or diamagnetic) depends on their nature (real
or complex). The relation between these scale parameters and coefficients of
QB equation (\ref{QB6}) is $k_{1}=\mu _{1}\mu _{2}\mu _{3}\mu _{4}$, $%
k_{2}=\mu _{1}\mu _{2}\mu _{3}+\mu _{1}\mu _{2}\mu _{4}+\mu _{1}\mu _{3}\mu
_{4}+\mu _{2}\mu _{3}\mu _{4}$, $k_{3}=\mu _{1}\mu _{2}+\mu _{2}\mu _{3}+\mu
_{1}\mu _{3}+\mu _{1}\mu _{4}+\mu _{2}\mu _{4}+\mu _{3}\mu _{4}$ and $%
k_{4}=\mu _{1}+\mu _{2}+\mu _{3}+\mu _{4}$. It is evident from above
equations that the eigenvalues are roots of the following quartic equation%
\begin{equation}
\mu ^{4}-k_{4}\mu ^{3}+k_{3}\mu ^{2}-k_{2}\mu +k_{1}=0.  \label{QBEV6}
\end{equation}%
It is critical to remember that plasma is considered to be incompressible in
this study, all information about fields and flows in the relaxed state is
contained within these scale parameters. The roots of above quartic equation
(\ref{QBEV6}) can be calculated using these relations 
\begin{eqnarray}
\mu _{1,2} &=&\frac{k_{4}}{4}+\frac{S\pm T}{2}, \\
\mu _{3,4} &=&\frac{k_{4}}{4}-\frac{S\pm L}{2},
\end{eqnarray}
where $S=0.5$($k_{4}^{2}-4k_{3}+4R$)$^{1/2}$, $R=$($\eta
_{3}k_{3}+3k_{2}k_{4}-12k_{1}-$($%
3k_{2}^{2}+3k_{1}k_{4}^{2}-12k_{1}k_{3}-k_{3}^{2}$)$^{2}-k_{3}^{2}$)$\eta
_{3}^{-1}$, $\eta _{1}=$($%
9k_{2}k_{3}k_{4}-36k_{1}-2k_{3}^{3}-k_{2}^{2}-27k_{1}k_{4}^{2}+108k_{1}k_{3}$
)/$54$, $\eta _{2}=$($3k_{2}k_{4}-12k_{1}-k_{3}^{2}$)/$9$ and $\eta _{3}=9$%
(( $\eta _{1}^{2}+\eta _{2}^{3}$)$^{1/2}+\eta _{1}$)$^{1/3}$. The $T$ and $L$
in case of $S\neq 0$ are given by $T=$($0.75k_{4}^{2}+0.25S^{-1}$($%
4k_{3}k_{4}-8k_{2}-k_{4}^{2}$)$-2k_{3}-S^{2}$)$^{1/2}$ and $L=$($%
0.75k_{4}^{2}-0.25S^{-1}$($4k_{3}k_{4}-8k_{2}-k_{4}^{2}$)$-2k_{3}-S^{2}$)$%
^{1/2}$ and in case of $S=0$, $T=$($2\sqrt{R^{2}-4k_{1}}+0.75k_{4}^{2}-2k_{3}
$)$^{1/2}$ and $L=$($0.75k_{4}^{2}-2\sqrt{R^{2}-4k_{1}}-2k_{3}$)$^{1/2}$.

From above relations it is clear that the scale parameters $\mu _{j}$ can be
real or complex and their values depend on plasma parameters. Figure (\ref%
{fig:1}) shows the character of scale parameters of quartic equation ($%
f(\mu)=\mu ^{4}-k_{4}\mu ^{3}+k_{3}\mu ^{2}-k_{2}\mu +k_{1}$) for slightly,
moderately and highly relativistic plasma species for given values of
Beltrami parameters and density ($a=2\mathbf{.}0$, $b=0\mathbf{.}8$, $c=3%
\mathbf{.}0$, $G_{i}=1.1$ and $N_{i}=0\mathbf{.}1$). In case of slightly
relativistic plasma species i.e. $G=1\mathbf{.}2$, the two eigenvalues are
real while other two are complex conjugate ($\mu _{1}=3\mathbf{.}0$, $\mu
_{2}=1\mathbf{.}6709$ and $\mu _{3,4}=0.5645\pm 0\mathbf{.}9889i$). By
increasing the temperature to $G=8\mathbf{.}0$, all the scale parameters
become real and given by $\mu _{1}=3\mathbf{.}0$, $\mu _{2}=1\mathbf{.}9386$%
, $\mu _{3}=0\mathbf{.}5642$ and $\mu _{4}=0\mathbf{.}291$. But in ultra-
relativistic regime $G=12\mathbf{.}0$, again all the scale parameters are
real ($\mu _{1}=3\mathbf{.}0$, $\mu _{2}=1\mathbf{.}9588$, $\mu _{3}=0%
\mathbf{.}6780$ and $\mu _{4}=0\mathbf{.}1631$). From plot it is also
evident that with an increase in $G$ the value of $\mu _{4}$ reduces whereas 
$\mu _{3}$ increases. It is important to highlight that such real scale
parameters give rise to paramagnetic behavior in the relaxed state. 
\begin{figure}[h]
\centering
\includegraphics{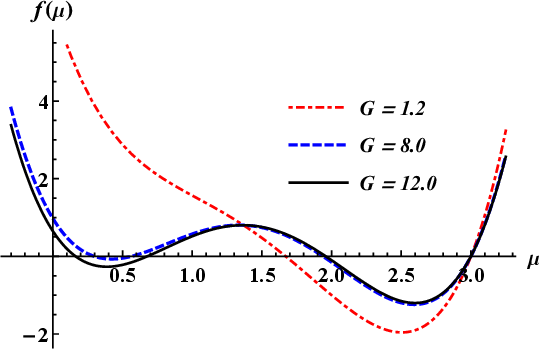}  
\caption{Character of scale parameters for different values of relativistic
temperatures when $a=2.0$, $b=0.8$, $c=3.0$, $G_{i}=1.1$ and $N_{i}=0.1$.}
\label{fig:1}
\end{figure}

\begin{figure}[h]
\centering
\includegraphics{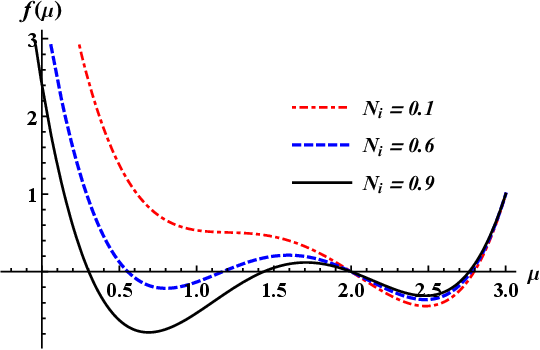}  
\caption{Character of scale parameters for different values of $N_{i}$ when $%
a=3.0$, $b=1.5$, $c=2.0$ and $G=1.5$ and $G_{i}=1.1$.}
\label{fig:2}
\end{figure}

Similarly figure (\ref{fig:2}) illustrates the character of scale parameters
for fixed values of Beltrami parameters and thermal energies of plasma
species ($a=3\mathbf{.}0$, $b=1\mathbf{.}5$, $c=2\mathbf{.}0$, $G=1.5$ and $%
G_{i}=1\mathbf{.}1$) for various values of ion density ($N_{i}=0\mathbf{.}1$%
, $0\mathbf{.}6$ \& $0\mathbf{.}9$). The plot demonstrates that for lower
ion density, two eigenvalues are real and other two are complex. For
instance when $N_{i}=0\mathbf{.}1,$ the eigenvalues are $\mu _{1}=2\mathbf{.}%
7954$, $\mu _{2}=1\mathbf{.}9999$ and $\mu _{3,4}=0\mathbf{.}8523\pm 0%
\mathbf{.}5246i $. Whereas for higher concentrations scale parameters are
real (when $N_{i}=0\mathbf{.}6$, the eigenvalues are real and given by $\mu
_{1}=2\mathbf{.}7767 $, $\mu _{2}=1\mathbf{.}9998$, $\mu _{3}=1\mathbf{.}1683
$ and $\mu _{4}=0\mathbf{.}5550$, and in case of $N_{i}=0\mathbf{.}9$, the
eigenvalues become $\mu _{1}=2\mathbf{.}7633$, $\mu _{2}=1\mathbf{.}9997$, $%
\mu _{3}=1\mathbf{.}4338$ and $\mu _{4}=0\mathbf{.}3031$). From the analysis
of plot it is clear that by increasing ion density the scale separation
between $\mu _{3}$ and $\mu _{4}$ increases. 
The occurrence of large and multiscale structures in hot astrophysical
plasmas, as well as their interpretation, has been a long-standing
challenge. Here, we'll look at a few different possibilities to the design
of system-size structures.

\begin{enumerate}
\item First we consider that $k_{1}=0$, then equation (\ref{QBEV6}) can be
written as $\mu $($\mu ^{3}-k_{4}\mu ^{2}+k_{3}\mu -k_{2}$)$=0$. The
preceding equation demonstrates that at least one system size structure ($%
\mu _{1}\sim 0$) exists in the relaxed state for this specific case. This is
also quantitatively verifiable. For example, when $b=1.7$, $c=2\mathbf{.}0$, 
$N_{i}=0\mathbf{.}1$, $G=7.0$ and $G_{i}=1.1$, the values of Beltrami
parameter $a$ and scale parameters are given by $a\approx -1.8883$, $%
\left\vert \mu _{1}\right\vert =1.8\times 10^{-17}$, $\left\vert \mu
_{2}\right\vert =1.809$, $\left\vert \mu _{3}\right\vert =1.6227$ and $%
\left\vert \mu _{4}\right\vert =2.0$. These scale parameters indicate that
for $k_{1}=0$, $\mu _{1}\approx 0$, $\mu _{2}\approx a$, $\mu _{3}\approx b$
and $\mu _{3}\approx c$.

\item Consider an other possibility for the creation of two system size
vortices i.e. $k_{1}=k_{2}=0$. Then equation (\ref{QBEV6}) can be written in
the following form $\mu ^{2}$($\mu ^{2}-k_{4}\mu +k_{3}$)$=0$, which leads
to following values of scale parameters $\mu _{1,2}=0$ and $\mu _{3,4}=0.5$($%
k_{4}\pm \sqrt{k_{4}^{2}-4k_{3}}$). To confirm this fact we consider the
following values of plasma parameters: $c=2\mathbf{.}0$, $N_{i}=0\mathbf{.}1$%
, $G=7.0$ and $G_{i}=1.1$, the values of Beltrami parameters ($a $, $b$) and
the scale parameters are given by $a\approx -0\mathbf{.}5492$, $b\approx
0.4943$, $\left\vert \mu _{1,2}\right\vert =9.34\times 10^{-9}$, $\left\vert
\mu _{3}\right\vert =2.0$ and $\left\vert \mu _{4}\right\vert =0.05488$.
These eigenvalues show that for $k_{1}=k_{2}=0$, $\mu _{1,2}\approx 0$ and $%
\mu _{3}\approx c$. This vast scale separation is important in the context
of dynamo mechanisms in plasmas.

\item For three macro scale structures we consider $k_{1}=k_{2}=k_{3}=0$,
then the quartic equation can be written as $\mu ^{3}$($\mu -k_{4}$)$=0$,
which yields the following eigenvalues: $\mu _{1,2,3}=0$ and $\mu _{4}=k_{4}%
\mathbf{.}$ For this case we consider $N_{i}=0\mathbf{.}1$, $G=7.0$ and $%
G_{i}=1.1$, the values of Beltrami parameters ($a$, $b$, $c$) and the scale
parameters are given by $a\approx 05536$, $b\approx -0\mathbf{.}4967$, $%
c\approx 0.0621$, $\left\vert \mu _{1,2,3}\right\vert \approx 1.58\times
10^{-6}$ and $\left\vert \mu _{4}\right\vert =0.2$.

\item When the ratios of generalized vorticities to the flows for pair
species are approximately equal i.e. $a\approx b$, the values of  two scale
parameters are $\mu _{1}=a$ and $\mu _{2}=c$.

\item When $a\approx b\approx c$, in this particular case two of roots are
equal i.e. $\mu _{1}\approx \mu _{2}\approx a$.
\end{enumerate}

\section{Analytical solution of QB state}, \label{s4}

The magnetic fields and coupled flows in the QB state are controlled by four
scale parameters. The size and nature of these scale parameters depends on
Beltrami parameters, relativistic temperatures and density of the plasma
species. To illustrate the effect of plasma parameters on the formation and
nature of QB self-organized structures, we will use a simple slab geometry
for an EPI plasma system. The general solution of QB equation (\ref{QB6})
can be represented as the linear combination of four distinct Beltrami
fields \cite{Yoshida1990} and can be written as%
\begin{equation}
\mathbf{B=}\sum\limits_{j=1}^{4}C_{j}\mathbf{B}_{j},
\end{equation}%
where $\mathbf{B}_{j}$ satisfies the following Beltrami condition $\mathbf{%
\nabla }\times \mathbf{B}_{j}=\mu _{j}\mathbf{B}$. In a simple slab
geometry, the magnetic field can be expressed as%
\begin{equation}
\mathbf{B=}\sum\limits_{j=1}^{4}C_{j}[\sin \left( \mu _{j}x\right) \widehat{y%
}+\cos \left( \mu _{j}x\right) \widehat{z}],  \label{SQB6}
\end{equation}%
where $C_{j}$ are constants and their values are given in appendix A. The
expression for composite flow velocity demands us to write its analytical
solution in the following manner:%
\begin{equation}
\mathbf{V=}\sum\limits_{j=1}^{4}F_{j}[\sin \left( \mu _{j}x\right) \widehat{y%
}+\cos \left( \mu _{j}x\right) \widehat{z}],  \label{SCV6}
\end{equation}%
where $F_{j}=C_{j}\left( k_{4}\mu _{j}^{3}-k_{3}\mu _{j}^{2}+k_{2}\mu
_{j}-k_{1}\right) $. The existence After formulating the analytical solution
for the QB magnetic field and associated flow, we may experiment with
different arbitrary values of plasma parameters to determine how they affect
relaxed state structures. For our current analysis of relaxed state
structures, we consider pulsar magnetospheric plasma. At a distance of $%
10^{8}$ cm from the pulsar surface, the electron density is $n_{e}=10^{6}$ cm%
$^{-3}$ and the corresponding skin depth is $5.31\times10^{2}$ cm \cite%
{Karpman1975,Melrose1978,Michel1982,Lazarus2012}.
\subsection{Impact of relativistic temperature}
\begin{figure}[h]
\centering
\includegraphics{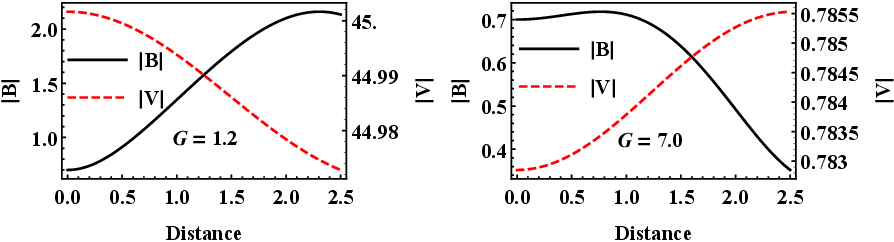}  
\caption{Trend of magnetic field (left vertical axis) and flow (right
vertical axis) for different values of relativistic temperature $G$ (1.2 \&
7.0) when $a=2.9$, $b=1.3$, $c=1.5$, $G_{i}=1.1$ and $N_{i}=0.1$.}
\label{fig:3}
\end{figure}

Besides the Beltrami parameters, the thermal energy and density of plasma
species have an important role in determining the magnitudes and nature of
the scale parameters. As mentioned earlier, the impact of ion relativistic
temperature is not significant due to large mass difference between pair and
ion species. So, in order to investigate the effect of thermal energy of
pair species, figure (\ref{fig:3}) shows the trend of magnetic field and
flow when the values of Beltrami parameters, relativistic ion temperature
and ion density are $a=2\mathbf{.}9$, $b=1\mathbf{.}3$, $c=1\mathbf{.}5$, $%
G_{i}=1\mathbf{.}1$ and $N_{i}=0\mathbf{.}1$, respectively. When the plasma
species are slightly relativistic such that $G=1.2$, the two eigenvalues are
real ($\mu _{1}=1\mathbf{.}5$, $\mu _{2}=2\mathbf{.}6395$) and other two are
complex conjugate ($\mu _{3,4}=0\mathbf{.}7802\pm 0\mathbf{.}7910i$). For
these eigenvalues, the magnetic field shows diamagnetic behavior while bulk
fluid velocity of plasma slightly decreases as strength of magnetic field
increases. In the case of highly relativistic electrons and positrons i.e. $%
G=7\mathbf{.}0$, all the scale parameters are real and distinct and their
values are $\mu _{1}=0\mathbf{.}1656$, $\mu _{2}=1\mathbf{.}1830$, $\mu
_{3}=1\mathbf{.}5$ and $\mu _{4}=2\mathbf{.}8513$. These all real
eigenvalues are consequence of higher thermal energies of pair species.
Corresponding to real eigenvalues, the magnetic field structure shows
paramagnetic behavior while flow has slightly increased as the magnetic
field decreases. From figure (\ref{fig:3}), it is evident that for certain
values of Beltrami parameters and ion density, an increase in thermal
energies of plasma species transforms the diamagnetic structures into
paramagnetic structures \cite{Mahajan1998}. 

\subsection{Impact of ion density}

\begin{figure}[h]
\centering
\includegraphics{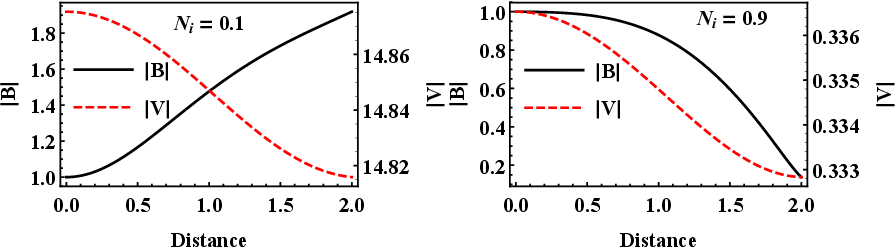}  
\caption{Trend of magnetic field (left vertical axis) and flow (right
vertical axis) for different values of ion density $N_{i}$ (0.1 \& 0.9) when 
$a=1.1$, $b=0.8$, $c=2.5$, $G=7.0$ and $G_{i}=1.1$.}
\label{fig:4}
\end{figure}

Figure (\ref{fig:4}) illustrates the effect of ion density on magnetic field
and flow structures for given thermal energies and Beltrami parameters ($%
G=7.0$, $G_{i}=1.1$, $a=1\mathbf{.}1$, $b=0\mathbf{.}8$ and $c=2\mathbf{.}5$%
). For this set of plasma parameters when $N_{i}=0\mathbf{.}1$, two of
eigenvalues are real while other two are complex and theses are $\mu _{1}=2%
\mathbf{.}4999$, $\mu _{2}=1\mathbf{.}0118$ and $\mu _{3,4}=0\mathbf{.}%
4441\pm 0\mathbf{.}2356i$. For these eigenvalues, figure (\ref{fig:4}) shows
diamagnetic behavior whereas flow trend is opposite to magnetic field. For
higher ion density $N_{i}=0\mathbf{.}9$, the scale parameters are real and
distinct, and have the following values: $\mu _{1}=2\mathbf{.}4998$, $\mu
_{2}=0\mathbf{.}9641,$ $\mu _{3}=0\mathbf{.}7579$ and $\mu _{4}=0\mathbf{.}%
1781$. For these real eigenvalues, Fig. (\ref{fig:4}) depicts that trend of
magnetic field is paramagnetic and bulk fluid velocity also decreases.\ From
above discussion, it can be concluded that for suitable Beltrami parameters,
the ion species density and relativistic temperature have a key role in
controlling the relaxed state structures. At higher relativistic
temperatures and ion densities, the magnetic structures are paramagnetic
while at lower values diamagnetic structures are formed. 

\subsection{Role of scale separation on field and flow generation}

The aforementioned study manifests a significant mutual coupling between the
magnetic field and velocity in thermally relativistic EPI plasma. This
interaction produces four self-organized structures corresponding to four
scale parameters ($\mu _{1}$, $\mu _{2}$, $\mu _{3}$ and $\mu _{4}$). When
the scale parameters are separated significantly, the magnetic field and the
coupled velocity may have an appreciable difference. Now we investigate how
disparate variations in the magnetic field and velocity are influenced by
the eigenvalues that are significantly disparate from one another (some of $%
\left\vert \mu _{j}\right\vert \gg 1$ and $\left\vert \mu _{j}\right\vert
\ll 1$) in the context of dynamo mechanisms. The dynamo process is a
fascinating subject in contemporary plasma physics, where magnetic fields
are self-generated by a moving electrically conducting fluid. In this
phenomenon, kinetic energy is transformed into magnetic energy \cite%
{Vainshtein1972,Moffatt1978}. The reverse of this phenomenon, where magnetic
energy is transformed to kinetic energy (generation of flow from a field),
can also occur and is termed as reverse dynamo \cite{Mahajan2005}. Although
the dynamo and reverse dynamo mechanisms are time-dependent but these
relaxed state fields and flows are the energy reservoir and provide the
seeds for these mechanisms.

\begin{figure}[h]
\centering
\includegraphics{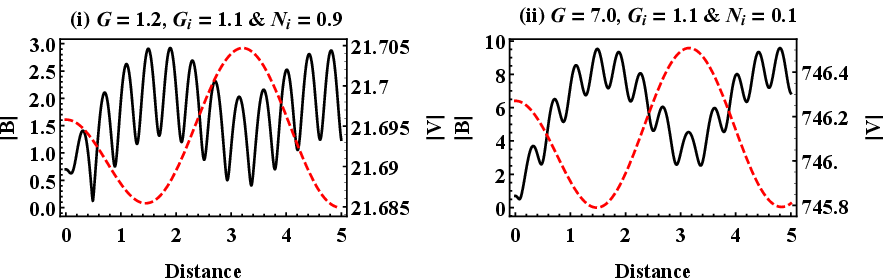}  
\caption{Trend of magnetic field (jittery (solid) -- left vertical axis) and flow
(smooth (dashed) -- right vertical axis) for different values of ion density and
relativistic temperatures when $a=17.0$, $b=16.0$ and $c=15.0$.}
\label{fig:5}
\end{figure}

To see the glimpse of field and flow generation mechanisms in QB state, we
consider two different regimes of Beltrami parameters while relativistic
temperature of pair species and ion density values are varied. First, we
assume that the values of Beltrami parameters are $a=17.0$, $b=16.0$ and $%
c=15.0$ and relativistic temperature of ion species is $G_{i}=1.1$. For this
set of Beltrami parameters all the scale parameters are real and distinct.
When ion density is high and pair species are mildly relativistic i.e., $%
N_{i}=0.9$, $G=1.2$ and $G_{i}=1.1$, the eigenvalues are $\mu _{1}\approx a$%
, $\mu _{2}\approx b$, $\mu _{3}\approx c$ and $\mu _{4}=0.05$. For these
eigenvalues, figure (\ref{fig:5}-i) clearly demonstrates that the variation
in the magnetic field is jittery, whereas the corresponding flow is smooth
and strong. For highly relativistic plasma ($G=7$) with small number of ions
($N_{i}=0.1$), the scale parameters have the following values: $\mu
_{1}\approx a$, $\mu _{2}\approx b$, $\mu _{3}\approx c$ and $\mu _{4}=0.016$%
. Figure (\ref{fig:5}-ii) illustrates that the strength of the magnetic
field and flow is drastically increased, despite the fact that the magnetic
field continues to be jittery. The decrease in ion density as well as the
increase in the effective masses of lighter species are the causes of this
increase in flow and field. The plot depicts a smooth flow beside a
fluctuating magnetic field.

Following the discussion presented above, it should be obvious that the flow
is more powerful than the magnetic field ($V\gg B$) when the scale
parameters reflect three microscale structures and one macroscale structure (%
$\mu _{1,2,3}\gg 1$ and $\mu _{4}\ll 1$). From plots, one can also deduce
that the small value of the magnetic field is being generated by the
conversion of kinetic energy into magnetic energy. During this energy
conversion process, the flow field is working against the Lorentz force \cite%
{Moffatt1985}. So the presence of three microscales and one macroscale
structure in QB state provide the seeds of a dynamo mechanism due to the
dominance of kinetic energy over magnetic energy in the equilibrium state.
The impact of scale separation in the relaxed states on the time-dependent
dynamo mechanism has been extensively investigated in recent years \cite%
{Mininni2002,Mininni2003,Mininni2005,Lingam2015}.

\begin{figure}[h]
\centering
\includegraphics{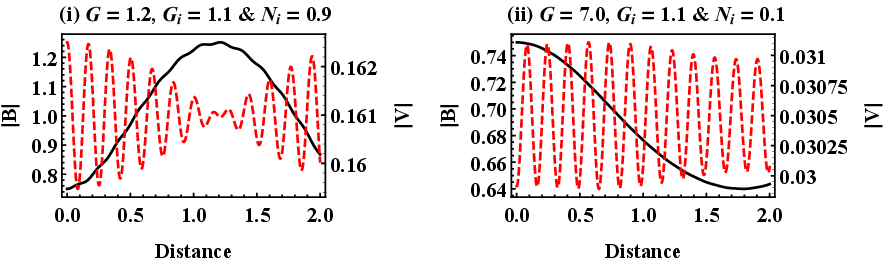}  
\caption{Trend of magnetic field (smooth (solid) -- left vertical axis) and flow
(jittery (dashed) -- right vertical axis) for different values of ion density and
relativistic temperatures when $a=2.0$, $b=4.0$, and $c=40.0$.}
\label{fig:6}
\end{figure}

In the second case, we consider $a=2.0$, $b=4.0$, $c=40.0$ and $G_{i}=1.1$.
For a slightly relativistic ($G=1.2$) plasma with higher ion density ($%
N_{i}=0.9$), the scale parameters are $\mu _{1}=0.6355$, $\mu _{2}=1.3834$, $%
\mu _{3}\approx b$ and $\mu _{4}\approx c$. Corresponding to these scale
parameters, figure (\ref{fig:6}-i) demonstrates that the magnetic field is
strong and smooth, but the flow is weak and jittery. In case of highly
relativistic ($G=7.0$) with a lower ion density ($N_{i}=0.1$), the scale
parameters have the following values: $\mu _{1}=0\mathbf{.}10$, $\mu
_{2}\approx a$, $\mu _{3}\approx b$ and $\mu _{4}\approx c$. For this set of
plasma parameters, figure (\ref{fig:6}-ii) shows that the strength of the
field and flow is decreased, despite the flow continuing to be jittery. This
decrease in field and flow is due variation in scale separation caused by
plasma parameters.

According to the description above and using equations (\ref{SQB6}-\ref{SCV6}%
) the relation between flow and field for these eigenvalues ($\mu _{1}\ll 1$%
, $\mu _{2,3}\sim 1$ and $\mu _{4}\gg 1$) shows that $V\ll B$ i.e., magnetic
energy is dominant over kinetic energy. It is also very clear from figure (%
\ref{fig:6}) that variation in magnetic field is smooth but the accompanying
flow is jittery and weak in comparison to the magnetic field. It seems that
the Lorentz force is driving the plasma flow and thus converts the magnetic
energy into kinetic energy. This conversion of magnetic energy into kinetic
energy happens along with the viscous heating which is the consequence of positive
work due to Lorentz force \cite{Brandenburg2019}. In a study pertaining to the heating of
the solar corona, Mahajan et al. demonstrated that, due to scale separation in the relaxed state,
the coupled magnetic field and flow can vary on vastly different length scales. So when a smooth
magnetic field is coupled with a jittery and weak flow, magnetic energy is dissipated along with
the viscous heating of plasma \cite{Mahajan2001}. So it is the scale separation
in the relaxed state and the resulting jittery components in both vector fields
that drive energy conversions via different scenarios. For
this particular situation, we are able to draw the conclusion that the
provision of the seeds for a reverse dynamo process is made possible when
the magnetic energy in the equilibrium state is significantly larger than
the kinetic energy. Such relaxed states where magnetic energy dominates the
kinetic energy are also utilized to study time-dependent reverse dynamo
processes \cite{Mahajan2005,Lingam2015,Kotorashvili2020}. The most recent
study by Kotorashvili and Shatashvili~investigated the unified dynamo and
reverse dynamo mechanisms in a three-component plasma composed of static
positive ions and two species of electrons: relativistic hot and
relativistic degenerate \cite{Kotorashvili2022}. So the mechanisms outlined
here for the generation of fields and flows in a variety of astrophysical
settings are extremely plausible. 

\section{Summary}, \label{s5}

The relaxation of a three-component thermally relativistic EPI magnetized
plasma has been explored in this paper. It is assumed that all of the plasma
species (electrons, positrons, and mobile positive ions) are
relativistically hot and the relativistic temperatures of electrons and
positrons are the same whereas the temperature of ion species is different
in this model of EPI plasma. When the steady state solutions of the
vorticity evolution equations for the plasma species are coupled using
Ampere's law, the relaxed state is a QB state, which is combination of four
single and distinct Beltrami states. As a result, there are four scale
parameters that describe the relaxed state of this plasma system. The
analytical solution of the QB relaxed state in a simple slab
geometry is provided in order to highlight the effect of plasma parameters
on the magnetic field and flow structures. Following the study, it has been
discovered that when the ion density and relativistic temperature of
electron and positron species are higher for given values of Beltrami
parameters, all the scale parameters are real and magnetic structures
exhibit the paramagnetic trend. It is critical to emphasize that the
influence of ion relativistic temperature is negligible on the relaxed state
structures due to large mass difference between ion and pair species.

Additionally, it has been investigated that when three microscale structures
along with a large scale structure are formed in QB state, the magnetic
variation occurs at considerably shorter length scales as compared to flow.
This flow and field pattern is an indicator that a magnetic field is being
generated by the strong plasma flow. The field and flow trends are switched
when there is a microscale structure, two structures of the order of
electron skin depth and one large-scale structure. In the relaxed state,
this flow field exhibits variations on shorter length scales which indicates
the presence of turbulence. Such disparate variations in field and flow
can provide the seeds for dynamo and reverse dynamo processes.  These findings
lead us to the conclusion that QB states are of great importance and can be used
to better understand the creation of multiscale structures and related
phenomena in thermally relativistic EPI plasmas found in astrophysical
and laboratory plasmas.

\begin{acknowledgments}
The work of M. Iqbal is funded by Higher Education Commission (HEC),
Pakistan under project No. 20-9408/Punjab/NRPU/R\&D/HEC/2017-18. We are also
thankful to the anonymous reviewer, whose insightful comments and suggestions
helped us to improve the quality of  the work.
\end{acknowledgments}

\begin{appendix}
\section*{Appendix A}
The constants $C_{j}$ can be
calculated with the help of following boundary conditions: $\left\vert 
\mathbf{B}_{z}\right\vert _{x=0}=b_{1}$, $\left\vert \mathbf{B}%
_{y}\right\vert _{x=x_{0}}=b_{2}$, $\left\vert (\mathbf{\nabla \times B}%
)_{z}\right\vert _{x=0}=b_{3}$ and $\left\vert (\mathbf{\nabla }\times 
\mathbf{B)}_{y}\right\vert _{x=x_{0}}=b_{4}$, where $b_{1}$, $b_{2}$, $b_{3}$, $b_{4}$ and $x_{0}$ are
arbitrary and real valued constants. The boundary conditions lead to these equations $%
\sum\limits_{j=1}^{4}C_{j}=b_{1}$, $\sum\limits_{j=1}^{4}C_{j}\sin \left(
\mu _{j}x_{0}\right) =b_{2}$, $\sum\limits_{j=1}^{4}C_{j}\mu _{j}=b_{3}$ and 
$\sum\limits_{j=1}^{4}C_{j}\mu _{j}\sin \left( \mu _{j}x_{0}\right) =b_{4}$. By solving these boundary condition
equations simultaneously one can obtain the values of $C_{j}$ given by $%
C_{j}=K^{-1}R_{j}$ where%
\begin{gather*}
R_{1}=\left[ \left( b_{3}\mu _{2}+b_{1}\mu _{3}\mu _{4}\right) \sin \left(
\mu _{2}x_{0}\right) -\left( b_{4}\mu _{2}+b_{2}\mu _{3}\mu _{4}\right) %
\right] \left[ \sin \left( \mu _{4}x_{0}\right) -\sin \left( \mu
_{3}x_{0}\right) \right]  \\
+\left[ \left( b_{3}\mu _{3}+b_{1}\mu _{4}\mu _{2}\right) \sin \left( \mu
_{3}x_{0}\right) -\left( b_{4}\mu _{3}+b_{2}\mu _{4}\mu _{2}\right) \right] %
\left[ \sin \left( \mu _{2}x_{0}\right) -\sin \left( \mu _{4}x_{0}\right) %
\right]  \\
+\left[ \left( b_{3}\mu _{4}+b_{1}\mu _{3}\mu _{2}\right) \sin \left( \mu
_{4}x_{0}\right) -\left( b_{4}\mu _{4}+b_{2}\mu _{3}\mu _{2}\right) \right] %
\left[ \sin \left( \mu _{3}x_{0}\right) -\sin \left( \mu _{2}x_{0}\right) %
\right] ,
\end{gather*}%
\begin{gather*}
R_{2}=\left[ \left( b_{3}\mu _{1}+b_{1}\mu _{3}\mu _{4}\right) \sin \left(
\mu _{1}x_{0}\right) -\left( b_{4}\mu _{1}+b_{2}\mu _{3}\mu _{4}\right) %
\right] \left[ \sin \left( \mu _{3}x_{0}\right) -\sin \left( \mu
_{4}x_{0}\right) \right]  \\
+\left[ \left( b_{3}\mu _{3}+b_{1}\mu _{1}\mu _{4}\right) \sin \left( \mu
_{3}x_{0}\right) -\left( b_{4}\mu _{3}+b_{2}\mu _{1}\mu _{4}\right) \right] %
\left[ \sin \left( \mu _{4}x_{0}\right) -\sin \left( \mu _{1}x_{0}\right) %
\right]  \\
+\left[ \left( b_{3}\mu _{4}+b_{1}\mu _{1}\mu _{3}\right) \sin \left( \mu
_{4}x_{0}\right) -\left( b_{4}\mu _{4}+b_{2}\mu _{1}\mu _{3}\right) \right] %
\left[ \sin \left( \mu _{1}x_{0}\right) -\sin \left( \mu _{3}x_{0}\right) %
\right] ,
\end{gather*}%
\begin{gather*}
R_{3}=\left[ \left( b_{3}\mu _{1}+b_{1}\mu _{2}\mu _{4}\right) \sin \left(
\mu _{1}x_{0}\right) -\left( b_{4}\mu _{1}+b_{2}\mu _{2}\mu _{4}\right) %
\right] \left[ \sin \left( \mu _{4}x_{0}\right) -\sin \left( \mu
_{2}x_{0}\right) \right]  \\
+\left[ \left( b_{3}\mu _{2}+b_{1}\mu _{1}\mu _{4}\right) \sin \left( \mu
_{2}x_{0}\right) -\left( b_{4}\mu _{2}+b_{2}\mu _{1}\mu _{4}\right) \right] %
\left[ \sin \left( \mu _{1}x_{0}\right) -\sin \left( \mu _{4}x_{0}\right) %
\right]  \\
+\left[ \left( b_{3}\mu _{4}+b_{1}\mu _{1}\mu _{2}\right) \sin \left( \mu
_{4}x_{0}\right) -\left( b_{4}\mu _{4}+b_{2}\mu _{1}\mu _{2}\right) \right] %
\left[ \sin \left( \mu _{2}x_{0}\right) -\sin \left( \mu _{1}x_{0}\right) %
\right] ,
\end{gather*}%
\begin{gather*}
R_{4}=\left[ \left( b_{3}\mu _{1}+b_{1}\mu _{2}\mu _{3}\right) \sin \left(
\mu _{1}x_{0}\right) -\left( b_{4}\mu _{1}+b_{2}\mu _{2}\mu _{3}\right) %
\right] \left[ \sin \left( \mu _{2}x_{0}\right) -\sin \left( \mu
_{3}x_{0}\right) \right]  \\
+\left[ \left( b_{3}\mu _{2}+b_{1}\mu _{1}\mu _{3}\right) \sin \left( \mu
_{2}x_{0}\right) -\left( b_{4}\mu _{2}+b_{2}\mu _{1}\mu _{3}\right) \right] %
\left[ \sin \left( \mu _{3}x_{0}\right) -\sin \left( \mu _{1}x_{0}\right) %
\right]  \\
+\left[ \left( b_{3}\mu _{3}+b_{1}\mu _{1}\mu _{2}\right) \sin \left( \mu
_{3}x_{0}\right) -\left( b_{4}\mu _{3}+b_{2}\mu _{1}\mu _{2}\right) \right] %
\left[ \sin \left( \mu _{1}x_{0}\right) -\sin \left( \mu _{2}x_{0}\right) %
\right] ,
\end{gather*}%
\begin{gather*}
K=\left( \mu _{1}\mu _{2}+\mu _{3}\mu _{4}\right) \left[ \left( \sin \left(
\mu _{1}x_{0}\right) -\sin \left( \mu _{2}x_{0}\right) \right) \left( \sin
\left( \mu _{3}x_{0}\right) -\sin \left( \mu _{4}x_{0}\right) \right) \right]
\\
+\left( \mu _{1}\mu _{4}+\mu _{3}\mu _{2}\right) \left[ \left( \sin \left(
\mu _{1}x_{0}\right) -\sin \left( \mu _{4}x_{0}\right) \right) \left( \sin
\left( \mu _{2}x_{0}\right) -\sin \left( \mu _{3}x_{0}\right) \right) \right]
\\
+\left( \mu _{1}\mu _{3}+\mu _{4}\mu _{2}\right) \left[ \left( \sin \left(
\mu _{2}x_{0}\right) -\sin \left( \mu _{4}x_{0}\right) \right) \left( \sin
\left( \mu _{3}x_{0}\right) -\sin \left( \mu _{1}x_{0}\right) \right) \right].
\end{gather*}
In order to do a numerical analysis of the relaxed state, the following
values of the boundary conditions are used (for Figs. 3 and 5):  $b_{1}=0\mathbf{.}7$, $b_{2}=0%
\mathbf{.}25$, $b_{3}=0\mathbf{.}2$ and $b_{4}=0\mathbf{.}03$. Whereas, for Fig. 4: $b_{1}=1\mathbf{.}0$, $b_{2}=0%
\mathbf{.}025$, $b_{3}=0\mathbf{.}3$ and $b_{4}=0\mathbf{.}5$, while for Fig. 6, these values are $b_{1}=0\mathbf{.}75$, $b_{2}=0%
\mathbf{.}35$, $b_{3}=0\mathbf{.}2$ and $b_{4}=0\mathbf{.}03$.
\end{appendix}

\end{document}